\documentclass[epj,final]{svjour}
\usepackage{graphicx}
\usepackage{amsmath}
\def\be{\begin{eqnarray}}
\def\ee{\end{eqnarray}}
\def\bc{\begin{center}}
\def\ec{\end{center}}

\begin{document}

\title{Observables of non-equilibrium phase transition}
\author{Boris Tom\'a\v{s}ik\inst{1,2}, Martin Schulc\inst{2}, 
Ivan Melo\inst{1,3}, and  Renata Kope\v{c}n\'a\inst{2}}
\institute{Univerzita Mateja Bela, FPV, Tajovsk\'eho 40,
97401 Bansk\'a Bystrica, Slovakia
 \and \v{C}esk\'e vysok\'e u\v{c}en\'i technick\'e v Praze, FJFI, B\v{r}ehov\'a 7,
11519 Prague 1, Czech Republic
\and \v{Z}ilinsk\'a univerzita, Elektrotechnick\'a fakulta, Akademick\'a 1, 01026 \v{Z}ilina, Slovakia}
\date{}
\abstract{
A rapidly expanding fireball which undergoes first-order phase transition 
will supercool and proceed via spinodal decomposition. Hadrons are produced
from the individual fragments as well as the left-over matter filling the space between them. 
Emission from fragments should be visible in rapidity correlations, particularly 
of protons. In addition to that, even within narrow centrality classes, rapidity distributions will 
be fluctuating from one event to another in case of fragmentation. This can be 
identified with the help of Kolmogorov-Smirnov test. Finally, we present a method
which allows to sort events with varying rapidity distributions
in such a way, that events with similar rapidity histograms are grouped together. 
}
\PACS{
{25.75.Dw} 
{21.65.-f} 
}

\maketitle

\section{Introduction}

Experiments at NICA aim to explore the region of the phase diagram where highly compressed 
and excited matter may undergo a first-order phase transition. It is argued elsewhere
in this volume that such a phase transition in a  rapidly expanding system 
may bring it out of equilibrium and end up in its spinodal decomposition. 
Such a process then generates enhanced fluctuations in spatial distributions of 
the baryon density and the energy density. 

In this paper we focus on observables which could help to identify such processes. 

Before we explain various possible observables, we introduce DRAGON: the  Monte Carlo 
tool suited for generation of hadron distributions coming from a fragmented fireball
\cite{dragon}. 
Then, we report on an idea proposed in \cite{Pratt:1994ye,randrup} 
and further elaborated in \cite{schulc}:
clustering of baryons can be visible in rapidity correlations of protons. Further, we turn 
our attention to the whole rapidity distributions of produced hadrons and present an 
idea to search for nonstatistical differences between them
with the help of Kolmogorov-Smirnov test \cite{KStest}. 
Finally, we propose a novel treatment now being developed which also 
compares momentum distributions from individual events and sorts events according 
to their similarity with each other \cite{kopecna}.


\section{Monte Carlo hadron production from fragments}

In order to test the effects of fireball fragmentation into droplets it is 
useful to have Monte Carlo tool for the generation of artificial events with such features 
included. One possibility is to construct hydrodynamic models
which include such a  behaviour in the evolution
\cite{Steinheimer:2012gc,Herold:2013bi,Steinheimer:2013gla,Steinheimer:2013xxa}. 
They allow to link the resulting 
effects in fireball evolution with the underlying properties of the hot matter. 
On the other hand, they offer less freedom for systematic investigation 
of how the fragmentation is indeed seen in  data. Interesting questions 
of this kind are: what is the minimum size and abundance of fragments that can 
be seen? What  exactly is their influence on spectra, correlations, anisotropies, 
and femtoscopy? How are these observables influenced by the combination 
of droplet production and collective expansion? 

Such questions can be conveniently explored with the help of Monte Carlo generator
that uses a \emph{parametrization} of the phase-space distribution of hadron 
production. Such a tool has been developed in \cite{dragon} under the 
title DRAGON (DRoplet and hAdron Generator fOr Nuclear collisions). 
All studies presented here have been performed on events generated
with its help.

The bedding of the generator is the blast-wave model. The probability to 
emit a hadron in phase-space is described by the emission function
\begin{multline}
S(x,p)\, d^4x = \frac{g}{(2\pi)^3} \, m_t \cosh(y-\eta)\,  
\exp\left ( - \frac{p_\mu u^\mu}{T} \right )\\
\times \Theta (R - r)\, 
\exp\left ( -\frac{(\eta - \eta_0)^2}{2 \Delta\eta^2} \right )
\delta(\tau - \tau_0)\\ 
\times \tau\,d\tau\, d\eta\, r\,dr\, d\phi\, .
\label{e:S}
\end{multline}
It is formulated in Milne coordinates  $\tau = \sqrt{t^2 - z^2}$,
$\eta = ({1}/{2}) \ln ((t+z)/(t-z))$ and polar coordinates $r$, $\phi$ in 
the transverse plane.
Emission points are distributed uniformly in transverse direction within 
the radius $R$ and freeze-out occurs along the hypersurface given by 
constant $\tau = \tau_0$. Azimuthal anisotropy has not been used in 
studies presented here although the model includes such a possibility. 
There is collective longitudinal and transverse expansion parametrized 
by the velocity field
\begin{eqnarray}
\nonumber
u^\mu &=& ( \cosh\eta\, \cosh\eta_t, \cos\phi\, \sinh\eta_t,  \\
&& \phantom{\cosh\eta\, \cosh\eta_t} \sin\phi \sinh\eta_t,\,
\sinh\eta \cosh\eta_t)\\
\eta_t & = & \eta_t(r) = \sqrt{2}\rho_0 \frac{r}{R}\,  .
\end{eqnarray}
The fireball is locally thermalized with the temperature $T$.

A part of the hadrons, which can be specified in the model, is emitted from 
the  drop\-lets. The drop\-lets stem from the fragmentation of the 
same hypersurface as assumed in eq.~(\ref{e:S}). The actual picture is that 
when the fireball fragments, some free hadrons are born between the produced 
droplets. The volume of droplets is  distributed according to 
\cite{mishustin}
\begin{equation}
{\cal P}_V(V) = \frac{V}{b^2} e^{-V/b}\,  .
\end{equation}
The average volume of droplets is then $2b$. The minimal mass is practically set 
by the lightest hadron in simulation: usually the pion. The probability to emit  hadron 
from a droplet drops exponentially in droplet proper time $\tau_d$
\begin{equation}
{\cal P}_\tau(\tau_d) = \frac{1}{R_d} e^{-\tau_d/R_d}\,  ,
\end{equation}
where $R_d$ is the radius of the droplet. Momenta of hadrons from droplets 
are chosen from the Boltzmann distribution with the same temperature as bulk production. 
Currently, neither momentum nor charge conservation is taken into account in droplet decays, but an upgrade of the model including these effects is envisaged. 


DRAGON also includes production of hadrons from resonance decays. Baryons 
up to 2~GeV and mesons up to 1.5~GeV of mass are included. Chemical
composition is specified by chemical freeze-out temperature and chemical potentials 
for baryon number and strangeness. (Chemical potential for $I_3$ should also be 
introduced but is practically very small and thus neglected in the simulations.)


\section{Proton correlations}

Hadrons emitted from the same droplet will have similar velocities. 
This should be seen in their correlations \cite{Pratt:1994ye,randrup}. 
Protons appear best suited for 
such a study. Their mass is higher than that of most mesons, so their deflection 
from the velocity of the droplet due to thermal smearing 
will be less severe. Pions would have better 
statistics thanks to their high abundance, but their smearing due to thermal 
motion and resonance decays is too big. 

Correlation function can be measured as a  function of rapidity difference 
$\Delta y = y_1 - y_2$ or (better) of the relative rapidity
\begin{equation}
y_{12} = \ln \left [ \gamma_{12} + \sqrt{\gamma_{12}^2 -1} \right ]
\end{equation}
with $\gamma_{12} = p_1\cdot p_2 / m_1 m_2$.

The correlation function is conveniently sampled as 
\begin{equation}
C_{12}(y_{12}) = \frac{P_2(y_{12})}{P_{2,\mathrm{mixed}}(y_{12})}
\end{equation}
where $P_2(y_{12})$ is the probability  to observe a pair of protons with relative 
rapidity $y_{12}$. The reference distribution $P_{2,\mathrm{mixed}}(y_{12})$
in the denominator is obtained via the mixed events technique. 

It is instructive to first consider a simple model where the rapidities of 
droplets follow Gaussian distribution
\begin{equation}
\zeta(y_d) = \frac{1}{\sqrt{2\pi \xi^2}} \exp\left ( - \frac{(y_d - y_0)^2}{2\xi^2}\right )
\,  .
\end{equation}
Within the droplet $i$ which has  rapidity $y_i$,  rapidities of protons are also 
distributed according to Gaussian
\begin{equation}
\rho_{1,i}(y) = \frac{\nu_i}{\sqrt{2\pi \sigma^2}} 
\exp\left ( - \frac{(y-y_i)^2}{2\sigma^2} \right )\,  .
\end{equation}
This distribution is normalized to the number of protons from droplet $i$, 
which is denoted as $\nu_i$. 

The resulting correlation function in this simple model is \cite{randrup,schulc}
\begin{multline}
C(\Delta y)-1 = \frac{\xi \langle N_d \rangle \langle \nu (\nu -1) \rangle_M}{%
\langle N_d (N_d - 1) \rangle\langle \nu \rangle_M^2}
\sqrt{1 + \frac{\sigma^2}{\xi^2}} 
\\
\frac{1}{\sigma} 
\exp\left ( - \frac{\Delta y^2}{4\sigma^2 \left ( 1 + \frac{\sigma^2}{\xi^2}\right ) }
\right )
\end{multline}
where $\langle N_d \rangle$ is the average number of droplets in one event
and $\langle \cdots \rangle_M$ denotes averaging over various droplets. 
Naturally, the width of the correlation function depends on $\sigma^2$, 
as might have been expected.  However, it also depends on the width of the 
rapidity distribution of  droplets: through the factor $(1 +\sigma^2/\xi^2)$, 
growing $\xi^2$ leads to narrower proton correlation function. 

As an illustration relevant for NICA we generated sets of events with the 
help of DRAGON. On these samples we studied the influence of droplet size 
and the share of particles from droplets on the resulting correlation functions. 
It turns out that the relative rapidity $y_{12}$ yields better results, so we have mainly 
used this observable in our analyses. A more detailed study, though not with 
specific NICA fireball settings, can be found in \cite{schulc}.

DRAGON was set with Gaussian rapidity distribution with the width of 1. 
Within the rapidity acceptance window $-1<y<1$ there were about 1200 
hadrons; this number includes  all neutral stable hadrons. Momentum 
distribution has been set by the temperature of 120~MeV and the transverse 
velocity gradient $\eta_f = 0.4$. Chemical composition was according to 
$T_{ch} = 140$~MeV and $\mu_B = 413$~MeV. Recall that resonance 
decays are included in the model. The same kinetic temperature and 
chemical composition was assumed for the droplets. Total mass of 
each droplet is given by its size and the energy density 0.7~GeV/fm$^3$.
Transverse size of the fireball was set to 10~fm and the lifetime $\tau=9$~fm/$c$,
but these parameters have no influence on the presented results. Note that 
we have imposed acceptance cut in rapidity $-1<y<1$, so that we do not show 
results that would not be measurable due to limited acceptance. 

In order to see the effect of droplet formation on the correlation function we 
simulated one data set with no droplets and three sets which differ in droplet 
settings. We have sets with: 
$b = 50$~fm$^3$ and the fraction of 25\% of hadrons from droplets, 
$b = 20$~fm$^3$ and 50\%,
$b = 20$~fm$^3$ and 75\%. Recall that the mean droplet volume is 
$2b$. 

\begin{figure}
\begin{center}
\includegraphics[width=0.48\textwidth]{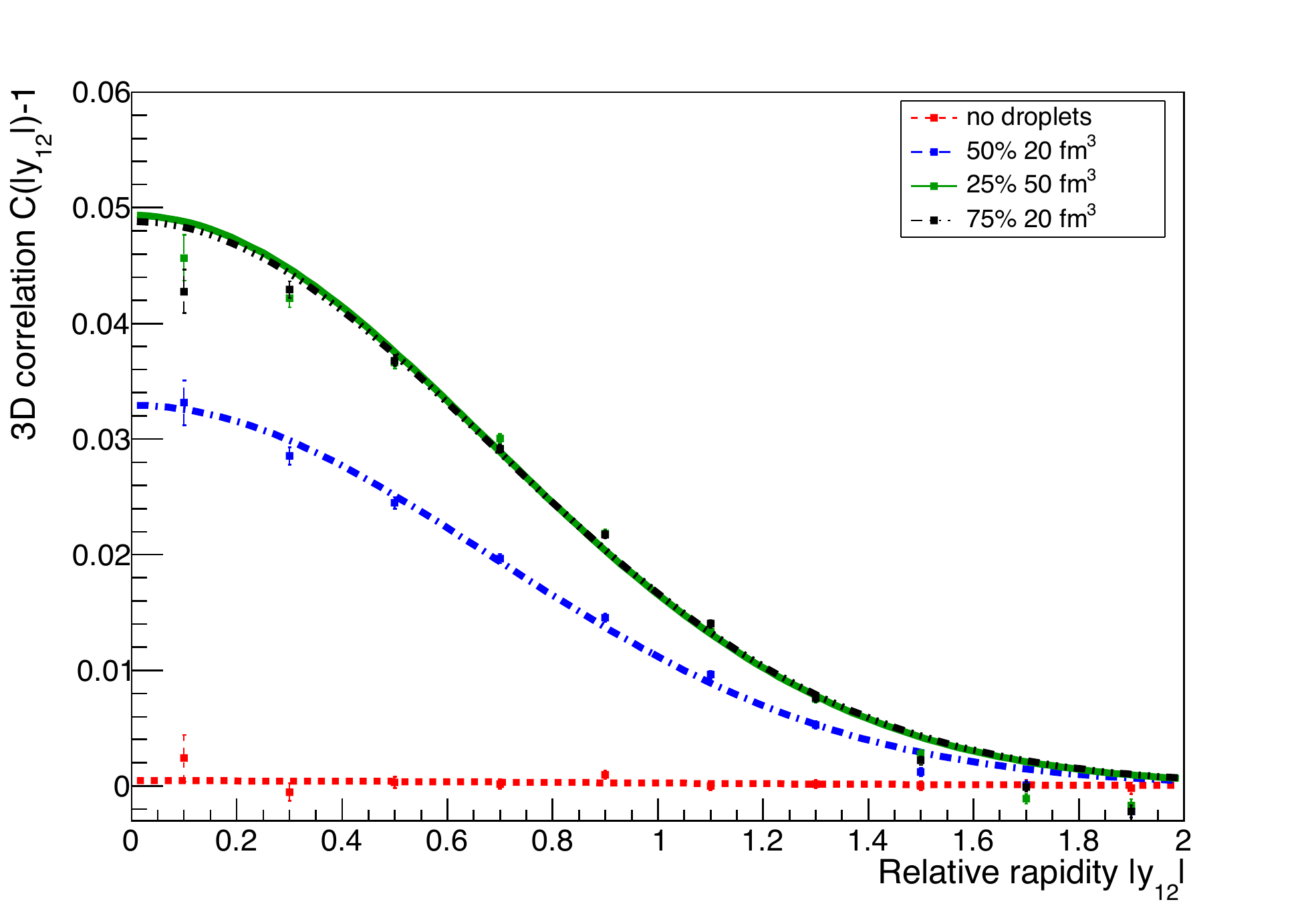}
\end{center}
\caption{Proton correlation functions for four different settings of hadron 
production from droplets.}
\label{f:pcorrel}
\end{figure}

The resulting proton correlation functions in $y_{12}$ are plotted in 
Fig.~\ref{f:pcorrel}. 
As expected, without fragmentation the correlation function is flat. The widths of 
the correlation functions are given by the smearing of the momenta of protons 
within one droplet, mainly due to temperature. 

The level of correlation is expressed in the height of the 
peak at $y_{12}=0$.  Naturally, this is expected to grow if a larger number of 
protons is correlated. This can be achieved in two ways: by increasing 
droplet sizes so that more protons come from each droplet, 
or by increasing the number of droplets by enhancing the share of particles 
produced by droplets. By coincidence we thus obtained very similar results 
for the cases with droplet fractions 25\% and 75\%, since the latter one assumes smaller 
droplets. 

Note the width scale of the correlation function which is larger than the 
typical scale of strong interactions. Thus any modification 
due to final state interactions which have not been included here 
is expected to be concentrated around the peak of our correlation 
functions.


\section{Comparison of rapidity distributions}

The fragmentation of the fireball actually leads to event-by-event 
fluctuations of rapidity distributions. In each event hadrons are 
produced from a different underlying rapidity distribution. In \cite{KStest}
it was proposed to use a standard statistical tool for the comparison of hadron
rapidity distributions from individual events: the Kolmogorov-Smirnov (KS) 
test. The KS test has been designed to answer the question, to what extent 
two empirical distributions seem to correspond to the same underlying probability 
density. 

\begin{figure}
\begin{center}
\includegraphics[width=0.49\textwidth]{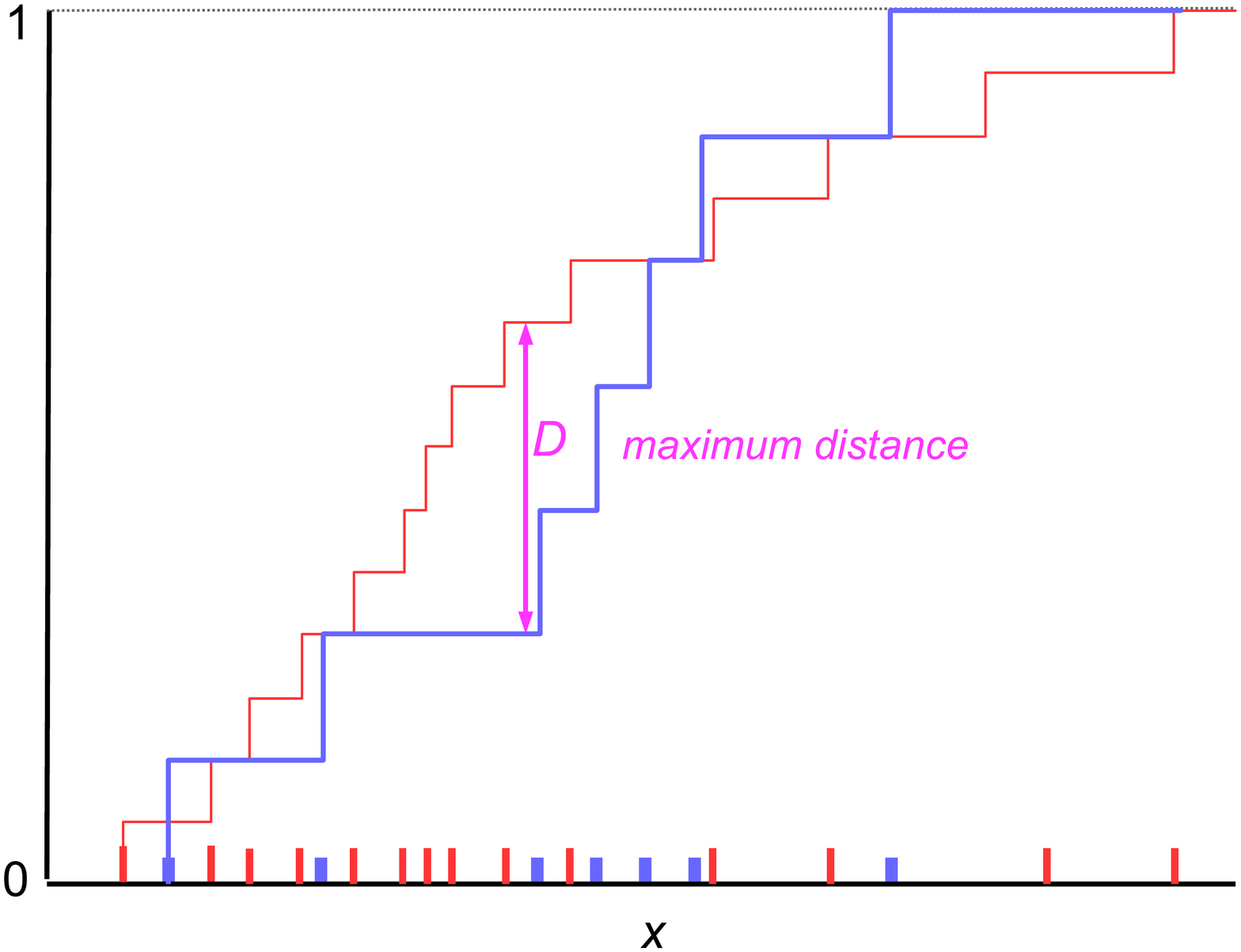}
\end{center}
\caption{Definition of the distance between two events. The measured values of 
variable $x$ are indicated on horizontal axis. Lines of different 
thickness  represent 
two different events.}
\label{f:KS}
\end{figure}

To apply the test on empirical distributions one first has to define a measure 
of how much they differ. For the sake of clarity and brevity we shall call 
empirical distributions \emph{events} and the measure of difference will 
be their scaled distance, to be defined later. A distance is defined in Fig.~\ref{f:KS}.
Consider measuring the quantity $x$ (this may be e.g.\ the rapidity) for
all particles in two different events. We mark the values of $x$ on the horizontal axis. 
Then, in the same plot we draw for each event its empirical cumulative distribution 
function. It is actually a staircase: we start at 0 and in each position where there is 
measured $x$ we make a step with the height $1/n_i$, where $n_i$ is the
multiplicity of the event. The maximum vertical distance $D$ between the two 
obtained staircases is taken as the measure of difference between the two events. 
For further work one takes the \emph{scaled distance}   
\begin{equation}
d = \sqrt{\frac{n_1 n_2}{n_1+n_2}} D
\end{equation}
where $n_1$, $n_2$ are the multiplicities of the two events. 

Next one defines 
\begin{equation}
Q(d) = P(d' > d)
\end{equation}
i.e.\ the probability that the scaled distance $d'$ determined for a pair of random 
events generated from the same underlying distribution will be bigger than $d$. 
The formulas for obtaining $Q(d)$ for any $d$ are given in the Appendix of 
\cite{KStest}. Thus defined, for large $d$, the value of $Q$ will be small because there 
is little chance that two events will be so much different. If all events come from 
the \emph{same} underlying distribution, then the $Q$'s determined on a 
large sample of event pairs will be distributed \emph{uniformly}. 

In a sample of events where the shape and dynamical state of the fireballs 
fluctuate, e.g.\ due to fragmentation, large scaled distance $d$ will be more
frequent. This is then translated into higher abundance of low $Q$ values. Thus 
non-statistical differences between events will show up as a peak at low $Q$
in the histogram of $Q$ values for large number of event pairs. In order to quantify the 
significance of the peak above the usual statistical fluctuations
we introduce
\begin{equation}
R = \frac{N_0 - \frac{N_{\mathrm{tot}}}{B}}{\sigma_0}
= \frac{N_0 - \frac{N_{\mathrm{tot}}}{B}}{\frac{N_{\mathrm{tot}}}{B}}
\end{equation}
where $N_0$ is the number of event pairs in the first $Q$-bin,
$N_{\mathrm{tot}}$ is the number of all event pairs and $B$ is the number 
of $Q$-bins. 

\begin{figure}
\begin{center}
\includegraphics[width=0.49\textwidth]{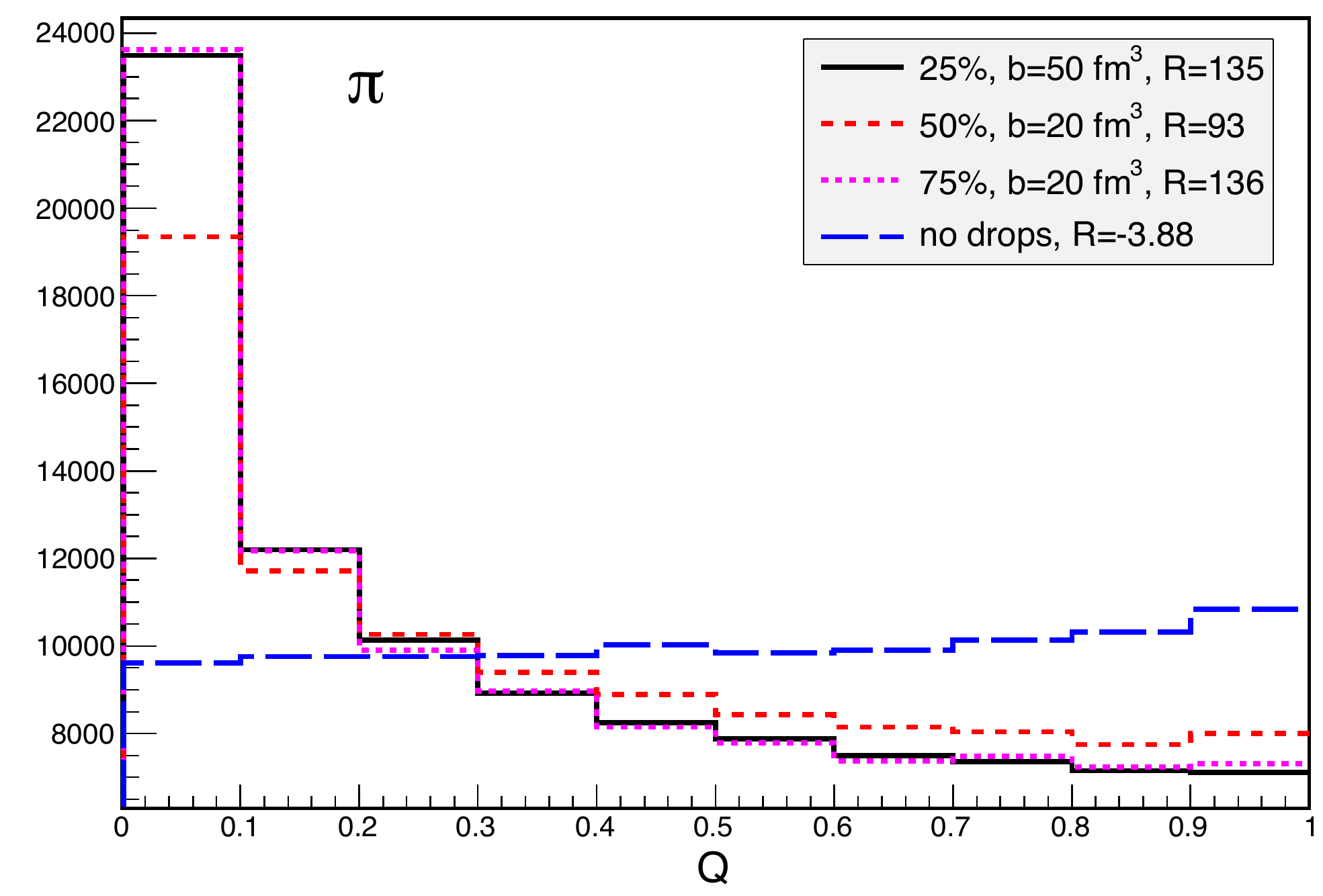}
\includegraphics[width=0.49\textwidth]{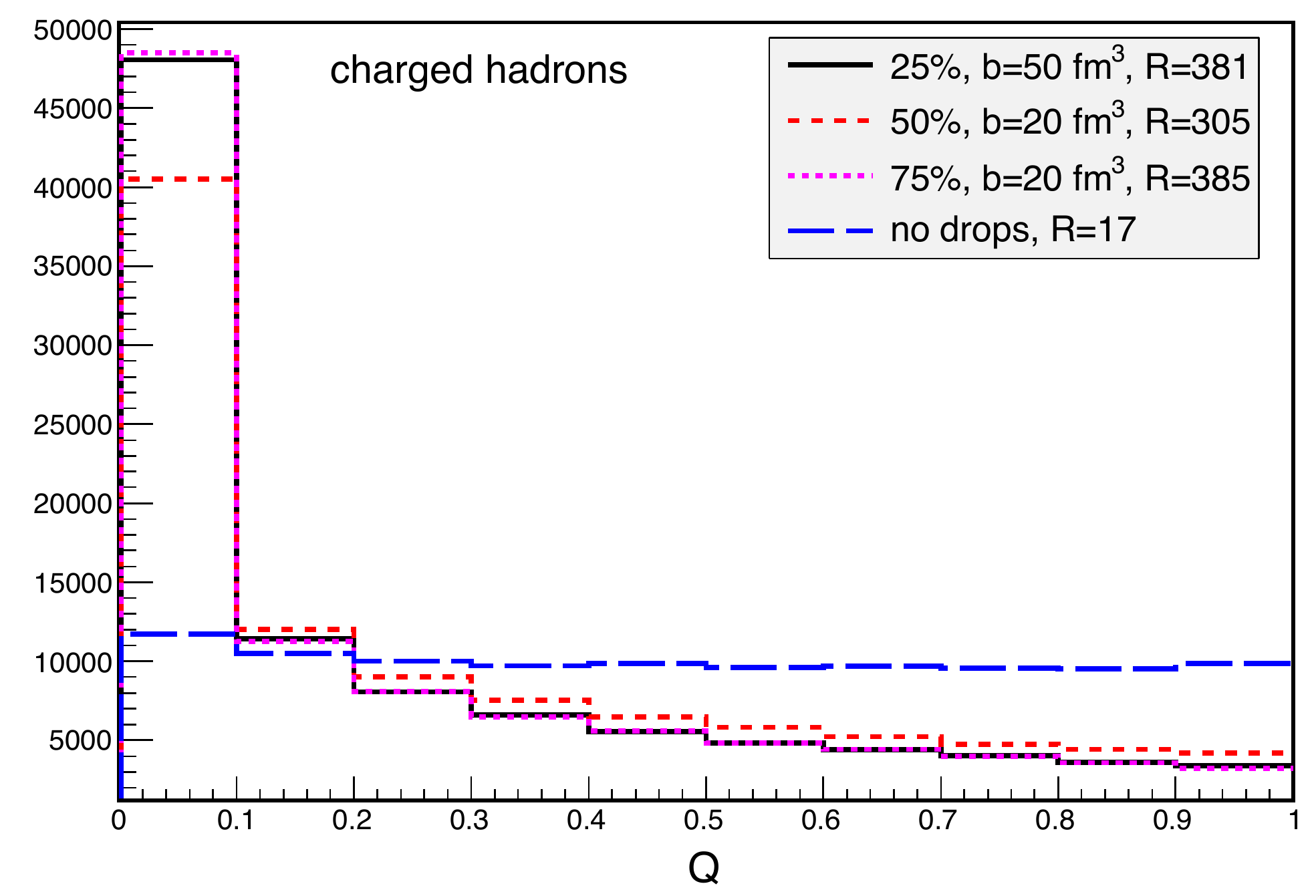}
\end{center}
\caption{$Q$-histograms for samples of $10^4$ simulated events. Rapidities 
of charged pions (top) and all charged hadrons (bottom) are taken into account.}
\label{f:Q}
\end{figure}

To illustrate the application at NICA, we have used event samples with the same 
settings as in the previous Section and show in Fig.~\ref{f:Q} the 
$Q$-histograms for pion rapidity distributions as well as rapidity distributions
of all charged hadrons. 
The signal is very strong and the one for charged hadrons is generally 
more pronounced than the one for pions. The comparison of different data sets 
is consistent with results for correlation functions from the previous section. 
Note that there is basically very weak signal for the case without droplets, which 
shows that clustering 
effect due to resonance decays cannot mask the investigated mechanism.


\section{Event shape sorting}

In presence of fireball fragmentation, rapidity distributions of different events show 
large variety. This motivates the quest to select among them groups of events 
which will be similar. Such groups allow to appreciate the range of fluctuations 
of the momentum distribution. They also may be useful for the construction 
of mixed events histograms used in correlation functions. 

A method for sorting events according to their similarity with each other has been 
proposed \cite{jackson,kopecna}. The application in \cite{kopecna} was on azimuthal 
angle distributions. Here we use it for rapidity distributions. Details can be 
found in \cite{kopecna}; here we only shortly explain the sorting algorithm.

An event is characterized when all its bin entries $n_i$ are given; $i$ numbers the 
bins in rapidity. Full bin record will be denoted $\{ n_i \}$. 

\begin{enumerate}
\item 
Events are initially sorted in a chosen way and divided into $N$ quantiles 
of the distribution. We use deciles, numbered by Greek letters.
\item 
For each event, characterized by record $\{ n_i \}$, calculate the 
probability that it belongs 
to the event bin $\mu$, $P(\mu|\{n_i\})$, using the Bayes' theorem
\begin{equation}
\label{e:Bay}
P(\mu|\{ n_i\} ) = \frac{P(\{ n_i\}| \mu ) P(\mu)}{P(\{n_i\})}\, .
\end{equation}
The probability $P(\{ n_i\}| \mu )$ that the event with bin record $\{n_i\}$ belongs 
to the event bin $\mu$ can be expressed as
\begin{equation}
P(\{ n_i\}| \mu ) = M! \prod_i \frac{P(i|\mu)^{n_i}}{n_i!}
\end{equation}
where $M$ is the event multiplicity, the product goes over all (rapidity) bins, and
$P(i|\mu)$ is the probability that a particle falls into bin $i$ in an event  from 
event bin $\mu$
\begin{equation}
P(i|\mu) = \frac{n_{\mu,i}}{M_\mu}\, .
\end{equation}
($M_\mu$ is the total multiplicity of all events in event bin $\mu$ and 
$n_{i,\mu}$ is the total number of particles in bin $i$.)
Coming back to eq.~(\ref{e:Bay}): $P(\mu)=1/N$ is the prior and 
\begin{equation}
P(\{n_i\}) = \sum_{\mu = 1}^N P(\{n_i\}|\mu) P(\mu)\,  .
\end{equation}
\item 
For each event determine
\begin{equation}
\bar \mu = \sum_{\mu = 1}^{N} P(\mu|\{n_i\}) \mu
\end{equation}
and re-sort all events according to $\bar\mu$. Then divide again into quantiles. 
\item 
If the ordering of events changed, re-iterate from point 2. In a less strict version 
of the algorithm, the ordering is re-iterated only if the assignment to quantiles has changed. 
\end{enumerate}
This iterative algorithm organizes events in such a way, that those which are 
similar to each other by the shapes of their histograms end up close 
to each other. It is not specified \emph{a priori}, however, whether there is any specific 
observable according to which the sorting proceeds. The algorithm itself picks 
the best ordering automatically. 
The method actually provides a more sophisticated version of the Event 
Shape Engineering.

\begin{figure}
\begin{center}
\includegraphics[width=0.49\textwidth]{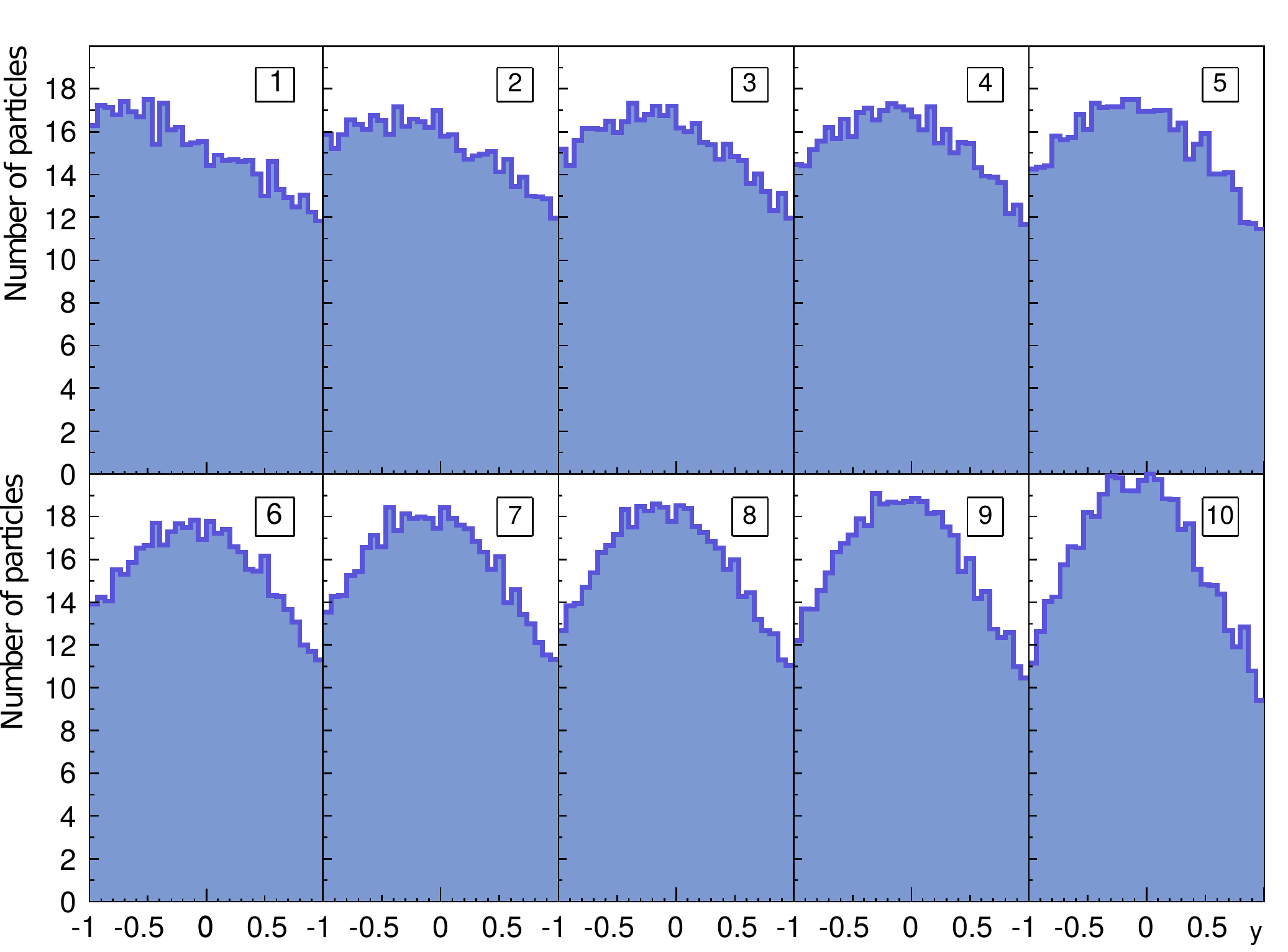}
\end{center}
\caption{Average rapidity histograms of the 10 event bins after the sorting algorithm 
with 5000 events
(with rapidity flip - see text) converged. Droplet fraction 25\% and $b = 50$~fm$^3$.}
\label{f:ess}
\end{figure}

We have tested the algorithm on a set of events generated by DRAGON with the same 
parameters as in previous two Sections. For illustration, we show in Fig.~\ref{f:ess}
the average histograms in different event bins after the sorting algorithm. 
We have chosen the data set with droplet fraction 25\% and $b=50$~fm$^3$
and the algorithm works with rapidity distributions of pions.
As a result of the fluctuations in rapidity distributions, the differences 
between event bins are large. On one end there are events with almost 
symmetric distributions, whereas on the other end there are events with strong 
emphasis on one side. 

It should be noted that the simulation setting assumes symmetric Gaussian rapidity 
distribution and corres\-pon\-ded to symmetric nuclear collisions. Consequently, there 
is no reason to favour one rapidity direction over the other. The resulting 
sorting in Fig.~\ref{f:ess} is obtained when in the middle of the iteration 
process one half of the events is flipped over the mid-rapidity. 

The difference between event bins is much bigger here than in a sample of events 
where no droplets are present.


\section{Conclusions}

We have sketched and explained two kinds of observables that can be used 
for identification of the fragmentation process: proton correlations in rapidity
\cite{randrup,schulc} and the 
Kol\-mo\-go\-rov-Smirnov test comparing the event-by-event
rapidity distributions \cite{KStest}. The motivation to look for the 
fragmentation comes
from the fact that a first order phase transition actually should proceed this way. 

It should be mentioned that in \cite{Torrieri:2007fb,Rajagopal:2009yw} it has been argued 
that potentially there is a mechanism which may lead to fireball fragmentation even in 
absence of the first order phase transition. A sharp peak of the \emph{bulk} viscosity 
as a function of temperature may suddenly cause resistance of  the bulk matter against  
expansion. Driven by the inertia, the fireball could choose to fragment. This possibility 
puts the uniqueness of the fragmentation process as the signature for the first order 
phase transition under question. Nevertheless, it is still certainly worthwhile to investigate 
the consequences of such a process. 

A process that could mask the signals of fragmentation is 
rescattering of hadrons emitted from droplets. It would be interesting to combine 
the presented methods with models including such a possibility.

Finally, we presented a method which is still being developed and which allows 
to sort the measured events automatically according to the most pronounced 
features in their histograms and build groups of similar events \cite{kopecna}. 
This would 
allow to study such groups, where event-by-event fluctuations are suppressed, in more
detail.


\acknowledgement

We acknowledge partial support  
by grants APVV-0050-11,
\linebreak VEGA 1/0469/15 (Slovakia). BT was also supported by 
grants RVO68407700,  LG15001 (Czech Republic).
RK acknowledges support from SGS15/093/OHK4/1T/14.


\end{document}